\documentclass{article}

\usepackage{arxiv}

\usepackage[utf8]{inputenc} 
\usepackage[T1]{fontenc}    
\usepackage{lmodern}        
\usepackage{hyperref}       
\usepackage{url}            
\usepackage{booktabs}       
\usepackage{amsfonts}       
\usepackage{nicefrac}       
\usepackage{microtype}      
\usepackage{graphicx}

\title{Social-ecological feedbacks drive tipping points in farming
system diversification}

\author{
    Melissa S. Chapman
   \\
    University of California, Berkeley \\
  Berkeley, California, USA \\
  \texttt{\href{mailto:mchapman@berkeley.edu}{\nolinkurl{mchapman@berkeley.edu}}} \\
   \And
    Serge Wiltshire
   \\
    University of California, Berkeley \\
  Berkeley, California, USA \\
  \texttt{} \\
   \And
    Patrick Baur
   \\
    University of Rhode Island \\
  Kingston, Rhode Island, USA \\
  \texttt{} \\
   \And
    Timothy Bowles
   \\
    University of California, Berkeley \\
  Berkeley, California, USA \\
  \texttt{} \\
   \And
    Liz Carlisle
   \\
    University of California, Santa Barbara \\
  Santa Barbara, California, USA \\
  \texttt{} \\
   \And
    Federico Castillo
   \\
    University of California, Berkeley \\
  Berkeley, California, USA \\
  \texttt{} \\
   \And
    Kenzo Esquivel
   \\
    University of California, Berkeley \\
  Berkeley, California, USA \\
  \texttt{} \\
   \And
    Sasha Gennet
   \\
    The Nature Conservancy \\
  Arlington, Virginia, USA \\
  \texttt{} \\
   \And
    Alastair Iles
   \\
    University of California, Berkeley \\
  Berkeley, California, USA \\
  \texttt{} \\
   \And
    Daniel Karp
   \\
    University of California, Davis \\
  Davis, California, USA \\
  \texttt{} \\
   \And
    Claire Kremen
   \\
    University of British Columbia \\
  Vancouver, BC, Canada \\
  \texttt{} \\
   \And
    Jeffrey Liebert
   \\
    Cornell University \\
  Ithaca, New York, USA \\
  \texttt{} \\
   \And
    Elissa M. Olimpi
   \\
    Virginia Tech \\
  Blacksburg, Virginia, USA \\
  \texttt{} \\
   \And
    Joanna Ory
   \\
    University of California, Berkeley \\
  Berkeley, California, USA \\
  \texttt{} \\
   \And
    Matthew Ryan
   \\
    Cornell University \\
  Ithaca, New York, USA \\
  \texttt{} \\
   \And
    Amber Sciligo
   \\
    The Organic Center \\
  Washington, DC, USA \\
  \texttt{} \\
   \And
    Jennifer Thompson
   \\
    University of California, Berkeley \\
  Berkeley, California, USA \\
  \texttt{} \\
   \And
    Hannah Waterhouse
   \\
    University of California, Berkeley \\
  Berkeley, California, USA \\
  \texttt{} \\
   \And
    Carl Boettiger
   \\
    University of California, Berkeley \\
  Berkeley, California, USA \\
  \texttt{\href{mailto:cboettig@berkeley.edu}{\nolinkurl{cboettig@berkeley.edu}}} \\
  }

\newlength{\cslhangindent}
\newlength{\csllabelwidth}
\newlength{\cslentryspacingunit} 
  {}%
  {\par}
\newenvironment{CSLReferences}[2] 
 {
  \setlength{\parindent}{0pt}
  \ifodd #1
  \let\oldpar\par
  \def\par{\hangindent=\cslhangindent\oldpar}
  \fi
  \setlength{\parskip}{#2\cslentryspacingunit}
 }%
 {}
\usepackage{calc}

\usepackage{amsmath}
\usepackage{amssymb}
\usepackage{lineno}
\usepackage{setspace}
\usepackage{booktabs}
\usepackage{longtable}
\usepackage{array}
\usepackage{multirow}
\usepackage{wrapfig}
\usepackage{float}
\usepackage{colortbl}
\usepackage{pdflscape}
\usepackage{tabu}
\usepackage{threeparttable}
\usepackage{threeparttablex}
\usepackage[normalem]{ulem}
\usepackage{makecell}
\usepackage{xcolor}
\begin{document}
\maketitle

\begin{abstract}
The emergence and impact of tipping points have garnered significant
interest in both the social and natural sciences. Despite widespread
recognition of the importance of feedbacks between human and natural
systems, it is often assumed that the observed nonlinear dynamics in
these coupled systems rests within either underlying human or natural
processes, rather than the rates at which they interact. Using adoption
of agricultural diversification practices as a case study, we show how
two stable management paradigms (one dominated by conventional,
homogeneous practices, the other by diversified practices) can emerge
purely from temporal feedbacks between human decisions and ecological
responses. We explore how this temporal mechanism of tipping points
provides insight into designing more effective interventions that
promote farmers' transitions towards sustainable agriculture. Moreover,
our flexible modeling framework could be applied to other cases to
provide insight into numerous questions in social-ecological systems
research and environmental policy.
\end{abstract}


\hypertarget{science-for-society}{%
\section{Science for Society}\label{science-for-society}}

Understanding the mechanisms of tipping points in social-ecological
systems is critical to designing effective policy interventions in
numerous environmental contexts. Using adoption of agricultural
diversification practices as a case study, we show how tipping points in
social-ecological systems can emerge purely from the temporal feedbacks
between human decisions and ecological responses. Further, we explore
why these finding matters for designing incentive programs to promote
farmers' transitions towards sustainable agriculture. We present a
flexible modeling framework that can be built on to address numerous
questions in social-ecological systems and environmental policy.

\hypertarget{introduction}{%
\section{Introduction}\label{introduction}}

Both ecosystems and social systems can change states abruptly as the
result of crossing critical thresholds. These critical thresholds
(``tipping points'', or states of a system where small perturbations can
trigger large responses) have garnered extensive academic and public
attention (Gladwell 2006; Rockström et al. 2009). However, mechanisms of
tipping points in social-ecological systems (SES) remain largely
explained by complex assumptions about either the ecological or social
system dynamics (Dai et al. 2012; Mumby, Hastings, and Edwards 2007;
Scheffer 2010; Horan et al. 2011b), rather than the ways in which these
systems interact.

In social-ecological systems, human actions impact ecological processes,
and the resultant ecological changes create feedbacks that alter future
management actions (Liu et al. 2007; Ostrom 2009; Walker et al. 2004).
These systems become challenging to model when the temporal dynamics of
ecological processes and their feedbacks to human systems (i.e.,
benefits from ecosystems services) do not align with the temporal scale
of human decision-making (Cumming, Cumming, and Redman 2006). Techniques
previously used to investigate both dynamic ecological processes and
decision-making in SES have mostly overlooked the temporal complexity of
decision-making (Lippe et al. 2019). For instance, agent-based models
are commonly used to explore complex emergent phenomena in SES. However,
these models often use single time-step, or user-defined, decision rules
rather than allowing for emergent decision strategies that maximize
expected rewards over longer time horizons (Lippe et al. 2019).
Similarly, economic models, which typically explicitly consider the time
horizons of decisions, often overlook ecological lags (J. H. Vandermeer
and Perfecto 2012). While temporal attributes are central drivers of
emergent dynamics in SES, social scientists have regularly pointed to
their importance for decision making processes as well (e.g., land
tenure affects decision-making by creating long-term incentives for
management (Fraser 2004; Long et al. 2017; Richardson Jr 2015; Soule,
Tegene, and Wiebe 2000)). New modeling approaches that can integrate
temporal attributes for both ecological change and human decision-making
are needed.

Agriculture is a particularly interesting case for exploring time lags
in social-ecological systems because many ecological responses to
management actions in these systems (such as planting hedgerows or
building up organic matter in soils) happen slowly, often taking years
to return ecological and/or financial benefits, which can exceed the
time frame of investment planning. Further, the duration of land tenure
varies considerably among farmers, which creates variation in, and
constraints on, horizons over which decisions strategies are optimized
(Soule, Tegene, and Wiebe 2000). Farmers on owned land may be able to
plan for payoffs that occur over the course of multiple decades or
generations. Tenant farmers who lease their farmland, by contrast, may
be constrained to the decisions that pay off during the length of lease
agreements. In the US, leases are most often short-term single-year
contracts but can extend up to 10 years (Bigelow, Borchers, and Hubbs
2016). Finally, while agriculture is regularly cited as a key driver of
anthropogenic ecological change (Foley et al. 2005, 2011; Stoate et al.
2009), different types of agriculture have radically different effects
on ecosystems. Some forms of agriculture rely on promoting ecological
processes that regenerate ecosystem services for their productivity and
are less input intensive (diversified farming systems), while others
rely primarily on external inputs, such as chemical fertilizers and
pesticides that often degrade the surrounding water, soil, and air
quality (Kremen, Iles, and Bacon 2012). In the context of diversified
farming systems, diversification practices include hedgerows, crop
rotation, intercrops, the use of compost, growing multiple crop types,
reduced tillage, and cover crops. This type of diversification is
distinct from the concept of operational diversification (i.e.,
increasing the range of revenue streams produced on a given farm, such
as tourism or value-added products) and has been shown to promote
ecosystem services that benefit farmers, including soil fertility and
water-holding capacity, pest and disease control, pollination and
productivity, thus providing an economically-viable alternative to
chemically-intensive methods of crop production (Tamburini et al. 2020;
Kremen and Miles 2012; Rosa-Schleich et al. 2019; Beillouin et al.
2021).

While adoption of diversified farm management practices encompasses a
continuum of actions and outcomes, suites of practices are often used
together in a package, coalescing around distinct stable states or
``syndromes'' (Andow and Hidaka 1989; Ong and Liao 2020; J. Vandermeer
1997a). The mechanisms used to explain and explore these patterns in
agricultural systems mathematically have relied on the assumption that
both ecological (or production) and decision (or economic) dynamics are
non-monotonic (a function that both increases and decreases) (J.
Vandermeer 1997a; J. H. Vandermeer and Perfecto 2012). In coupled
dynamic equations, if either of these systems is approximated as
monotonic (a function that only increases or only decreases), the larger
social-ecological system is characterized by a single stable point (or
no stable point), making multiple syndromes of production impossible to
explain with dynamic equations (J. Vandermeer 1997a; J. Vandermeer and
Perfecto 2012). In other words, the existence of distinct stable states
in agriculture -- defined by high levels of biodiversity and associated
ecosystem services on one hand and low-levels of biodiversity and
comparatively high synthetic inputs on the other -- cannot be explained
in conventional models without assuming complex structural dynamics.
While non-monotonic assumptions are often reasonable, equilibrium
explanations overlook the temporal component of both the ecological and
decision processes central to agricultural SES.

Markov Decision Processes (MDP) provide a convenient mathematical
framework for modeling decision making (Bellman 1957) in SES because
they allow for: (1) formulation of situations in which environments (in
this case, agroecosystems) change slowly and stochastically and (2) land
management decisions are forward looking and based on predictions about
how those decisions will impact a farmer's productivity and vitality in
the future. While MDPs have been widely used in a variety of
environmental control problems (Marescot et al. 2013), they are rarely
applied to modeling and exploring the dynamics of social-ecological
systems. Additionally, like other modeling approaches, these methods are
scarcely informed by, or ground truthed with, social science data.
Leveraging social science data, such as interviews or surveys, can help
inform critical features of social-ecological models.

This paper presents a stylized Markov Decision Process model of the
adoption of agricultural diversification practices to explore the
ecosystem service patterns that result specifically from interactions
between forward looking decision-making and a slowly-changing
environment(see Experimental Procedures for further details). Our
modeling work is inspired by patterns and system characteristics
(e.g.~the concept of forward-looking decision-making) that emerged from
the extensive empirical fieldwork with farmers that our research team
has conducted on commercial farms in California (see Experimental
Procedures for further details) (Gonthier et al. 2019; E. Olimpi et al.
2020; E. M. Olimpi et al. 2019) and through an iterative, collaborative
process with an interdisciplinary team comprising plant and soil
scientists, agricultural economists, ecologists, agricultural
sociologists, modelers, policy analysts, and farmers. Using this model,
we explore a mechanism leading to the two prevailing management
paradigms (i.e.~relying primarily on ecosystem services versus external
chemical inputs) that is the result not of complex structural
assumptions within either the human or ecological system, but rather the
rates at which the two systems interact. While our model necessarily
simplifies both decision-making and environmental processes, it provides
a useful framework to explore emergent properties in social-ecological
systems. We show that our findings have important implications, both for
agricultural policy implementation (such as incentive design) and
social-ecological systems theory.

\hypertarget{results}{%
\section{Results}\label{results}}

We observe the behavior of farmers' sequential choices and the resultant
environmental outcomes through time. The decision strategy, \(\pi^*\),
describes the emergent optimal course of action for a given ecosystem
service state (the stationary optimal state-dependent decision strategy;
see Experimental Procedures for further details). Figure 1A shows this
optimal strategy when the farmer plans over a discounted infinite time
horizon. Notably, it shows that at some ecosystem service state, the
optimal decision strategy displays a tipping point in which it becomes
advantageous to adopt diversification practices (Figure 1A).

We find that following the optimal decision strategy from Figure 1A,
farms have largely settled into two stable ecosystem service states,
with some farms transitioning to more simplified (lower levels of
ecosystem services) farming systems, and others to more diversified
(higher levels of ecosystem services) systems (Figures 1B and 1C).

\begin{figure}

{\centering \includegraphics{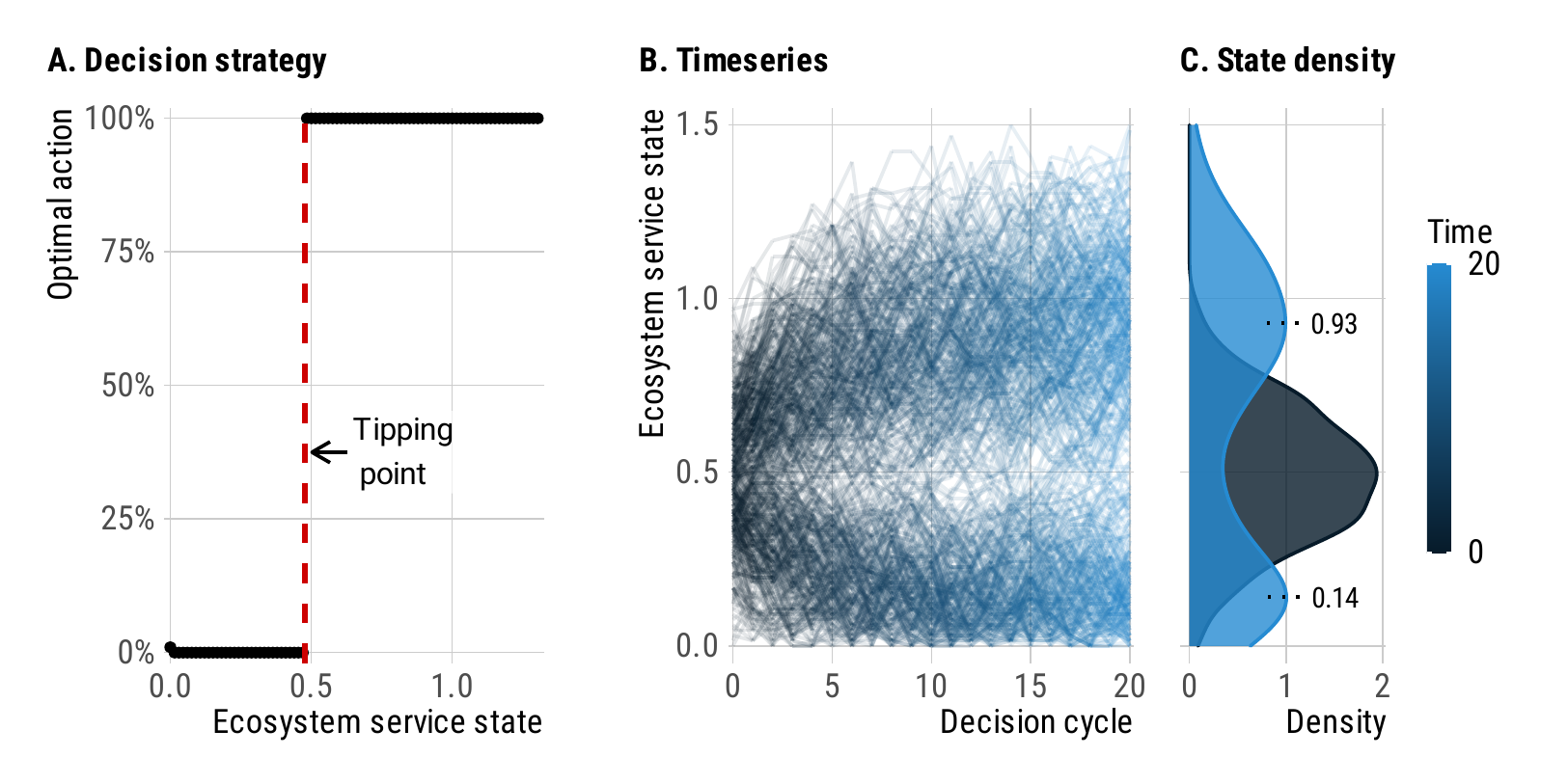} 

}

\caption{Initial ecosystem states (dark blue) are distributed normally (mean = 0.5; S.D. = 0.2; truncated at [0,1]).  (A) Optimal decision strategy $\pi$ for discounted infinite decision horizon. (B) Ecosystem state of each agent following decision strategy from (A) over 20 decision cycles (500 simulations).  (C) Initial ecosystem state density (dark blue) and final bimodal ecosystem state density at $t = 20$ (light blue). Density represents the probability density of a given ecosystem service state.}\label{fig:res_bimodal}
\end{figure}

\hypertarget{importance-of-temporal-dynamics-in-coupled-systems}{%
\subsection{Importance of temporal dynamics in coupled
systems}\label{importance-of-temporal-dynamics-in-coupled-systems}}

Our baseline model shows how a simple coupling of human choices and
ecological responses can result in bistable landscapes of high and low
diversification practice adoption and, as a result, high and low levels
of ecosystem services (Figure 1). By varying the time horizon of the
decision process, the rate of ecological response, and the cost/benefit
ratio, we find that this tipping point in decision strategy disappears
when the speed of response of either the ecological system or
decision-making process overwhelms the coupling (we use this as a proxy
for decoupling) (Figure 2A).

With temporal human/environment interactions, there exists a region of
cost-benefit ratio within which various decision tipping points and
bimodal ecosystem service state distributions exist, as in Figure 2A and
Figure 1). Intuitively, at low enough cost-benefit ratios, bimodality
disappears because farmers are expected to always invest (Figure 2A
bottom panel). Similarly, at high enough cost-benefit rations,
biomodality disappears because farmers are expected to always divest
(Figure 2A bottom panel). However, within a range of cost-benefit
ratios, decision strategies are expected to drive bimodal ecosystem
patterns (Figure 2A bottom panel between red dotted lines). Shortening
the time horizon of decisions (Fig 2B) or increasing the rate of
ecological processes (Fig 2C) necessarily changes the ratio of benefits
to costs required to make investing in practices worthwhile. However,
when decisions become temporally myopic (in this case, with a time
horizon of 2 decision cycles), the potential for bistability in adoption
trajectories disappears (Fig 2B bottom panel). Unlike Figure 1A, there
does not exist a region of cost-benefit space for this case in which
bistable patterns of ecosystem states exist (Figure 2B bottom panel).
Similarly, when ecological processes become fast enough that the
ecosystem responds almost immediately to farmer actions (\(r=0.95\)),
alternate stable states fail to emerge, regardless of cost-benefit
ratios (Figure 2C bottom panel). Only when both a gradually changing
environment and a forward-looking decision-maker (i.e.~a farmer who
takes into account potential benefits over the long term) are coupled,
do tipping point phenomena emerge in the decision strategy, leading to
two predominant ecosystem service states (Figure 2A, Figure 1). This
bimodal pattern matches farmers' experiences based on quotes from our
interview data (Figure 7; see Experimental Procedures for further
details), as well as other real world agricultural systems (J. H.
Vandermeer and Perfecto 2012).

\begin{figure}

{\centering \includegraphics{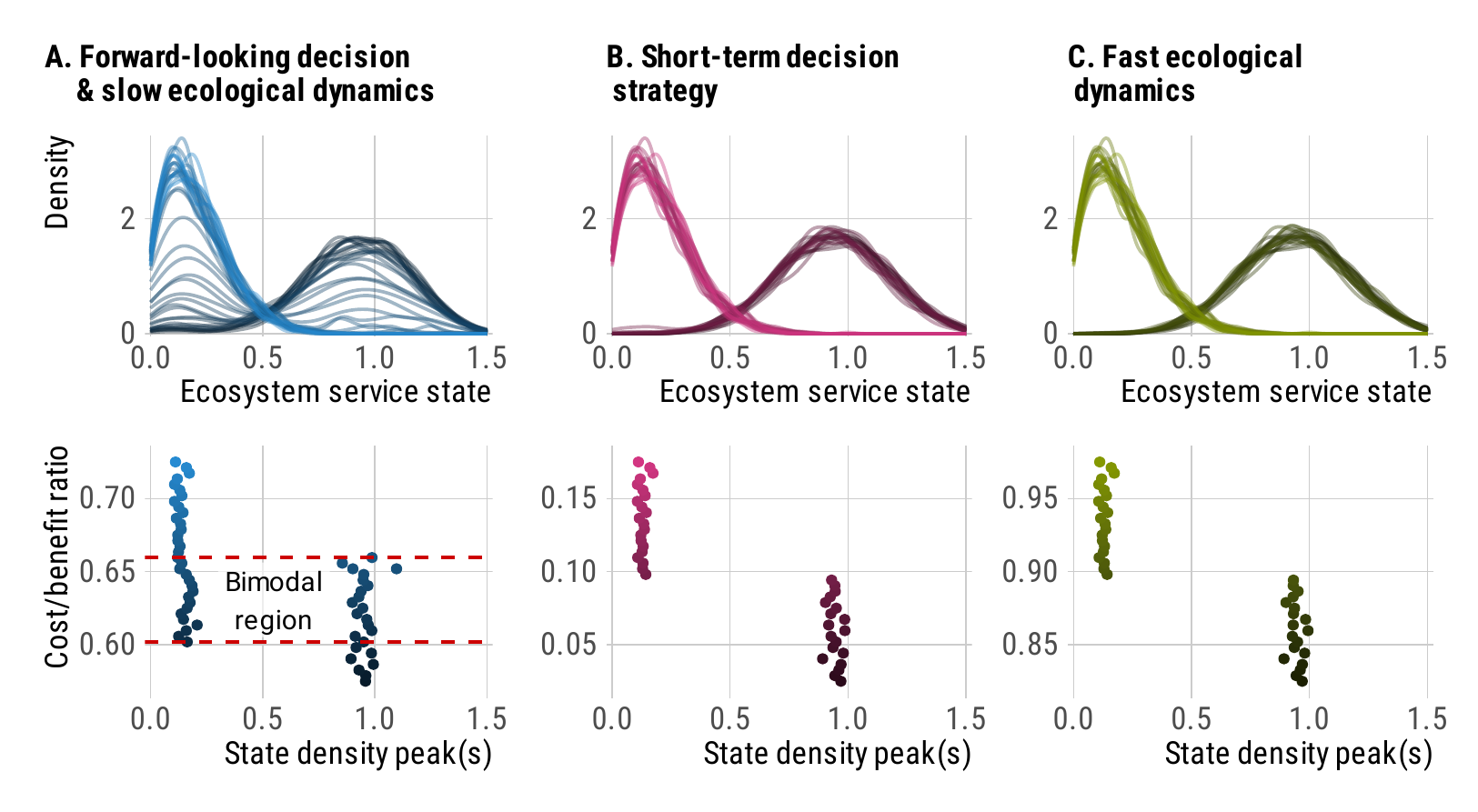} 

}

\caption{For three scenarios (coupled human/natural system, overly-myopic decision maker, and overly-fast ecological change), cost/benefit ratio was varied incrementally over 40 values, indicated by color shade, across a $c:b$ range of width 0.15, encompassing the transition between a 'never invest' to an 'always invest' policy.  For each $c:b$, 500 replicate simulations were conducted as in Fig 3.  Upper plots show distribution of final ecosystem service  state for each $c:b$.  Lower plots show density curve peak(s). Where overlap is observed in the lower graphs indicates the c:b ratios associated with bistability.  (A) By coupling a forward-looking decision-maker (e.g., a farmer who takes into account potential benefits over the long term) and a slowly-adapting environment, complex dynamics like alternate stable states can emerge (seen in cost benefit ratios between the red dotted lines). Bistable states do not exist at all cost-benefit ratios in this case (i.e., at a high enough cost to benefit ratio no adoption will occur, leading to a single low adoption state). Further, with (B) a short-term decision strategy (solving the MDP over a 2-year time horizon), or (C) a fast ecological change rate ($r = 0.95$), no bimodality is observed. In the cases of (B) and (C), the shift from no adoption to all-in adoption exists at some cost benefit ratio, removing the possibility of bistability in (A)}\label{fig:res_b_sweep}
\end{figure}

\hypertarget{influence-of-land-tenure-policy-on-ecosystem-service-states}{%
\subsection{Influence of land tenure policy on ecosystem service
states}\label{influence-of-land-tenure-policy-on-ecosystem-service-states}}

Given that temporal factors emerged as central themes from our interview
data on diversified farming adoption patterns (Figure 7), and that such
factors are more broadly relevant to understanding decision making
patterns on rented land (Soule, Tegene, and Wiebe 2000), we investigated
the impact of land tenure policy on farmer decision making.

We solved the MDP from Figure 1 on a constrained time horizon
(10-decision cycles, in comparison to an infinite time horizon in Fig
1), representing the shorter horizon on which tenant farmers might make
decisions.

Comparing the final state distribution of the long-tenure (baseline)
versus the short-tenure model shows that, as a farmer's expected land
tenure duration decreases, it becomes optimal to reduce diversification
adoption across a wider range of ecosystem states (Figure 3). This
results in ecosystem state degradation even among farm sites with an
initially high ecosystem service value, with all farms ending up in the
simplified state after 20 decision cycles (which might represent two
separate 10 year leases) (Figure 3C). However, the duration of land
tenure may not be the sole factor defining decision horizons. Numerous
economic and cultural factors -- for example, whether farmers are highly
motivated to seek sustainability as a goal in itself rather than solely
for individual economic reasons -- might also impact the time frame over
which a farmer is willing to wait for ecological benefit.

\begin{figure}

{\centering \includegraphics{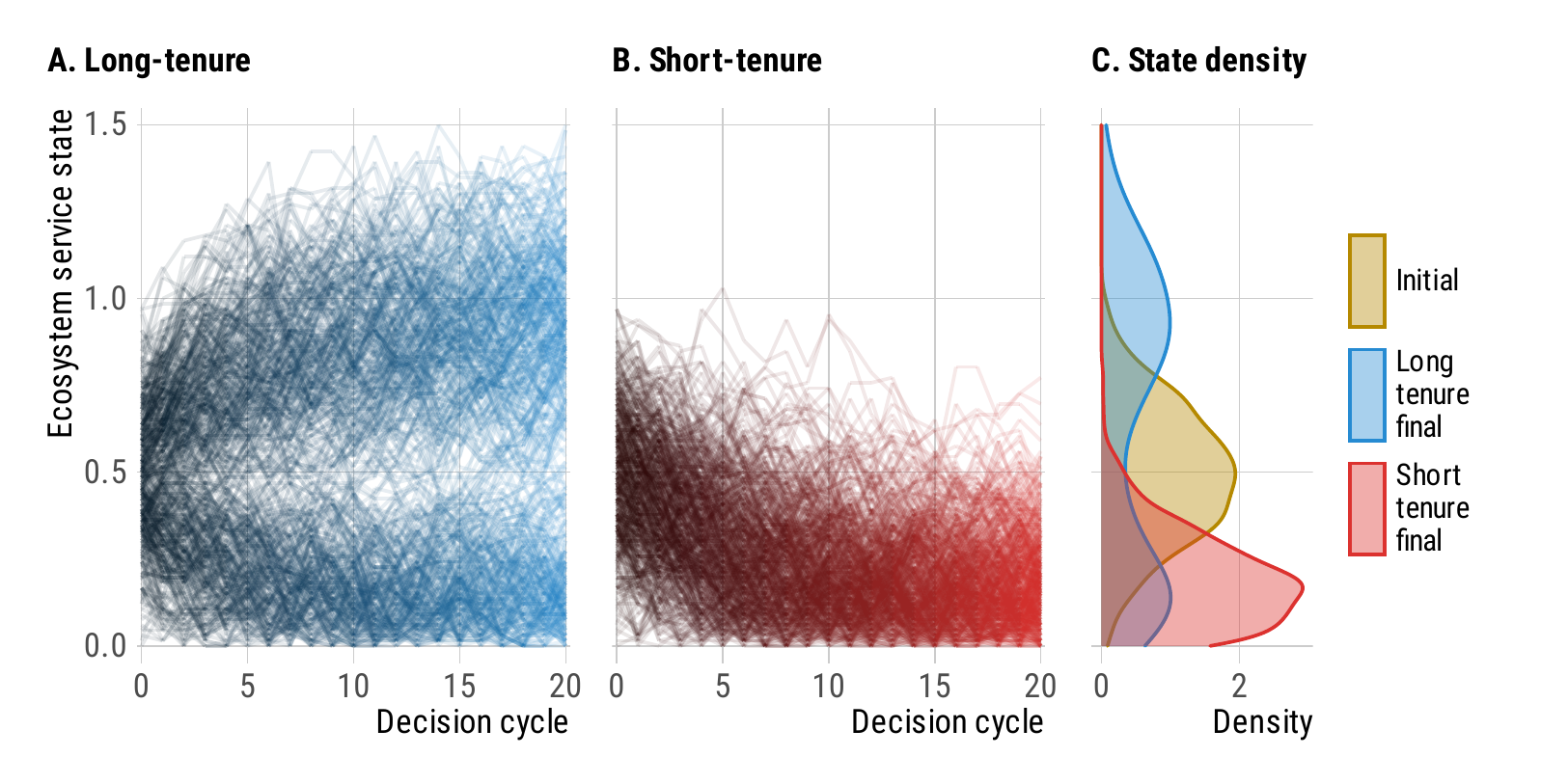} 

}

\caption{(A) The simulation is identical to that in Fig 3B, and represents long, stable land tenure. (B) The model from (A) is solved under a finite, 10-decision time horizon (rather than an infinite time horizon) to represent short-tenure. (C) Comparison between final state distribution of short- vs. long-tenure model runs.}\label{fig:res_tenure}
\end{figure}

\begin{figure}

{\centering \includegraphics{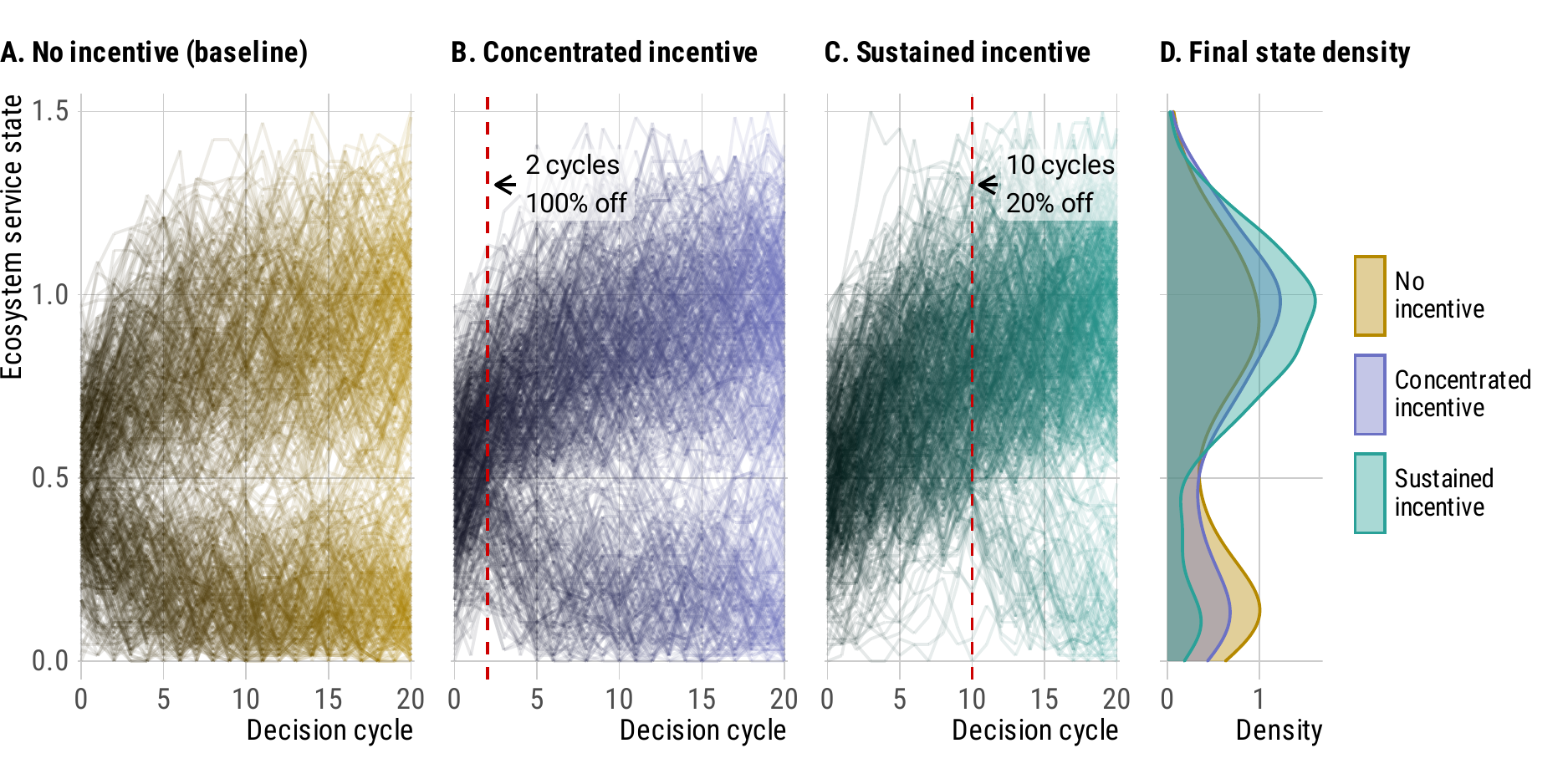} 

}

\caption{Starting from the same initial states as Fig 3, ecosystem service state time series are shown for (A) a large, abrupt incentive (100\% of adoption expenses are covered for two years) vs. (B) a smaller, more sustained incentive (i.e. adoption cost is 80\% of baseline for 10 years).  Before discounting, both packages have the same total cost to the funder (the equivalent of 2 years' worth of full adoption cost offsets). With discounting, (C) scenario is cheaper. After the incentive period, farmers (agents) adjust their decision rules to that of the base case (i.e. no incentive) until $t = 20$.  (D) Shows that the sustained incentive ultimately drove more DP adoption.}\label{fig:res_incentive}
\end{figure}

\hypertarget{temporal-dynamics-and-incentive-structures}{%
\subsection{Temporal dynamics and incentive
structures}\label{temporal-dynamics-and-incentive-structures}}

Our coupled social-ecological system model also allows for exploration
of how incentives that shift cost-benefit structures influence
management practices. Based on feedback from the farmers we interviewed
(Figure 7), we explore the impact of incentive duration on the efficacy
of policies to promote adoption of diversification practices by
comparing two different publicly funded incentive scheme designs: a
short-term (two-time step) incentive which fully covers the cost of
adoption, versus a longer-term (ten-time step) incentive that only
partially offsets the adoption cost. Both schemes offer the same total
amount of financial support. Within the model, farmers adapt their
optimal decision strategies for the given cost-benefit ratio during the
incentive period, and at its conclusion they revert to the baseline
strategy (i.e.~without payments).

We find longer, more sustained incentive programs to be more effective
at pushing the farmer over the critical threshold toward diversified
farming (Figure 4). Once a farmer has crossed the viable ecosystem
service state threshold (or optimal decision strategy tipping point), it
becomes less likely that they will return to simplified systems, even
after incentives are removed. Because it takes a series of investment
actions for the ecosystem service state to cross this threshold,
longer-term incentives ultimately result in more adoption of
diversification practices. Additionally, because the agent is
forward-looking, they are able to assess the entire expected reward of a
long term incentive.

\hypertarget{discussion}{%
\section{Discussion}\label{discussion}}

Our analysis suggests a mechanism for tipping points in
social-ecological systems that does not rely on complex assumptions
about the structure of the social or ecological systems alone. Instead,
these tipping points emerge from the temporal interactions between
forward-looking decisions (i.e., a farmer who considers potential
benefits over the long term) and slowly emerging ecological outcomes.
While alternate stable states within social-ecological systems,
including farming systems, have been previously explored and observed
(Horan et al. 2011a; J. Vandermeer 1997b; J. H. Vandermeer and Perfecto
2012), our results shed light specifically on temporal feedbacks that
might contribute to this pattern (Figure 2). We also show how path
dependence can result in self-perpetuating low ecosystem states and low
adoption of diversification practices (Figure 1) and why this provides
novel insights not only for social-ecological research (Figure 2), but
also for agricultural policy (Figure 3 and Figure 4).

In contrast to equilibrium models (J. H. Vandermeer and Perfecto 2012),
our model assumes (Experimental Procedures; Figure 6) that ecological
and environmental processes take time to respond to the adoption of a
diversified practice. For example, soil organic matter and its benefits
(such as improved water retention and storage of essential nutrients)
take years to build after starting practices like cover cropping and
compost additions (Poeplau and Don 2015). Our interviews with farmers
support this reality.\\
One farmer explains:

\begin{quote}
``I'll use five years, which seems like a long time, but I mean, that's
only potentially 5 or 10 crop cycles depending how heavy you
crop\ldots There's probably some very good soils that can be turned
around relatively quickly if everything works right. Somebody might see
some pretty dramatic benefits in a year or two, depending how bold they
wanted to do things. But I think the changes in soil in my mind, they're
not immediate. You don't make grand changes right away. So I mean, if
you get started doing some reduced tillage using more cover crops, if
you have a good source of compost and start incorporating those
practices, I would hope that you would see something in five years.''
\end{quote}

We show how time delays in ecosystem responses to management decisions,
as exemplified above, can explain patterns of multiple stable ecosystem
service states (Figure 5 P1). While existing explanations of multiple
stable states in SES provided by equilibrium models (J. H. Vandermeer
and Perfecto 2012) are not necessarily wrong, temporal explanations for
this pattern reflect key system attributes described by farmers (Figure
7) and allow for the exploration of intervention strategies that are
temporally constrained (e.g.~land tenure, incentives, etc.). While not
addressed in this analysis, the interaction of nonmonotonic (or
generally more complex) subsystem dynamics and the temporal interactions
of those subsystems will be an important path for future research.

\begin{figure}

{\centering \includegraphics[width=1\textwidth]{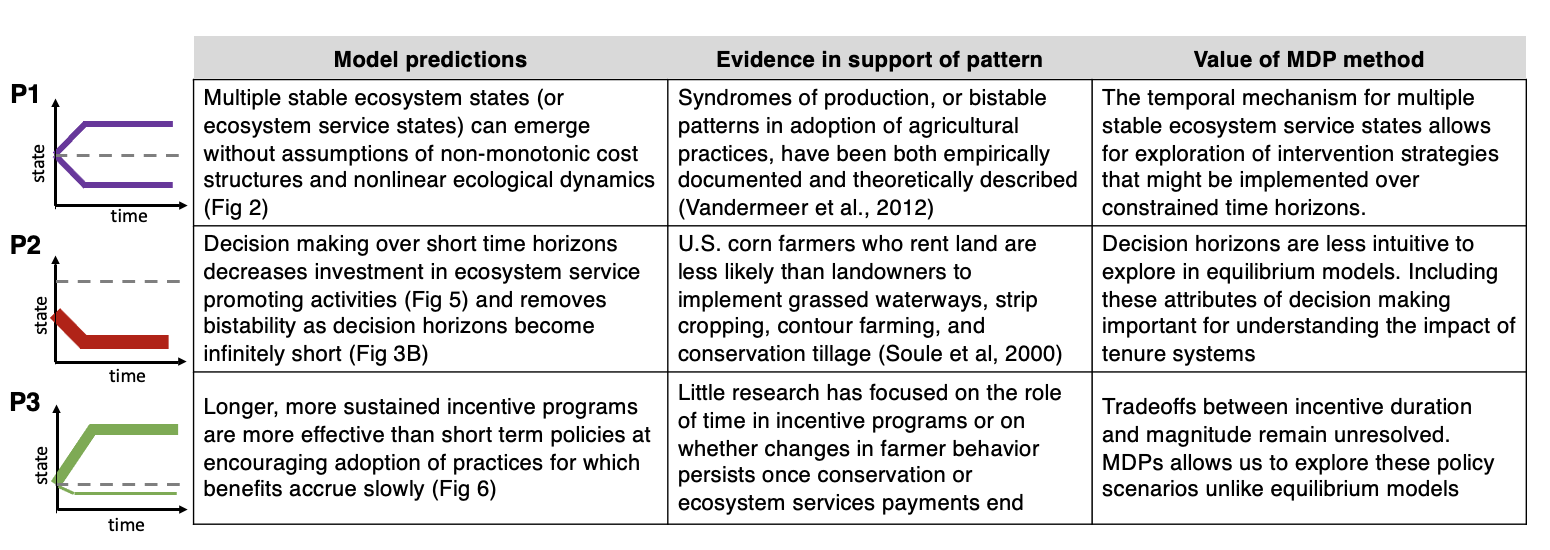} 

}

\caption{(Table 1) Table of the main model predictions, evidence in support of the pattern, value added of the temporal mechanism and minimal assumptions.}\label{fig:predictions}
\end{figure}

Our results also have important implications for understanding farmer
decision-making and agricultural policy design. Our model explains why
the land tenure status of a farmer can significantly influence their
willingness and ability to adopt diversification practices (Figure 1;
Figure 5 P2). This finding accords with a large body of sociological
research documenting how security and length of land tenure affect the
adoption of sustainable agricultural practices (Fraser 2004; Long et al.
2017; Richardson Jr 2015; Soule, Tegene, and Wiebe 2000), suggesting
that our model captures emergent socio-ecological dynamics of farming
systems. As another farmer explains, ``We do have hedgerows on several
of the ranches, more where we have long-term leases.'' Growers who hold
shorter leases are more likely to decide that adopting diversification
practices will not benefit them. They may lose their investment if their
lease ends forcibly or may have insufficient time to learn how to use
practices in the particular ecological and geographical conditions of
their farm (Calo and De Master 2016; Calo 2018). Immigrant farmers and
farmers of color, especially those new or beginning, often struggle to
achieve stable land tenure due to racial discrimination, poverty, or
language barriers in farmer networks, policy, and finance (Minkoff-Zern
2019).\\
Thus, policies which specifically aim to increase land tenure, for
example by supporting ownership and generational succession, may be
powerful levers to effect positive change in this area.

Finally, our model suggests that existing incentive programs to promote
agricultural sustainability and ecosystem services by reducing the costs
of practice adoption may need significant redesign (Figure 4; Figure 5
P3). Such policies have become an integral part of farming over the past
half-century (Batáry et al. 2015; Graddy-Lovelace and Diamond 2017).
They are particularly interesting to explore with a Markov Decision
Process due to their often sequential but time-limited nature. Incentive
policies rolled out over a given time frame are challenging to study
with equilibrium analyses or simple decision rules.

Our results suggest that long-term \textbf{sustained} incentives, even
when only partially covering the cost of adoption, may be more effective
in shifting farmers from simplified ecological states to diversified
states than more concentrated short-term incentives. We show that the
cost of interventions and the social-environmental benefit of those
interventions are not necessarily equivalent. Rather, the perceived
stability of incentive programs may be an important driver of adoption.
This dynamic can be overlooked when the temporal rates of coupled
dynamics in social-environmental systems are not considered. If farmers
expect a stable source of support over a known time period, they may
decide it is worthwhile to experiment and persist with a new practice
that may not provide observable benefits for many years (Claassen et al.
2014). Unstable support, by contrast, may lead to farmers abandoning
practices after a short time, or may prevent farmers from trying new
conservation practices (Dayer et al. 2018). Moreover, the reduced
transaction costs that come with farmers making a longer-term
commitment, while not captured in our model, would only further suggest
the higher efficacy of sustained incentives compared to concentrated
incentives.

This finding is particularly relevant to the design of government
payment programs and suggests that smaller payments can be highly
effective in encouraging the adoption of diversification practices (or
other ecosystem service promoting practices) when distributed over long
time horizons. Small payments over a longer time frame also constitute a
lower total cost to the government when considering even modest discount
rates. Yet, the relationship between the length of incentive programs
and the persistence of changes in land manager behavior once payments
end remains unclear. One study found that when landowners were unable to
re-enroll in a waterbird habitat program in northern California due to
three-year period limits, participant numbers declined and farmers
persisted less with their practices (Dayer et al. 2018). Other studies
have found that growers tend to switch back land that is left unused in
return for payments via the federal Conservation Reserve Program to
`more valuable' productive uses (e.g., corn ethanol (Roberts and
Lubowski 2007)). It is possible, as our model suggests, that steady, if
somewhat lower, conservation payments might result in more favorable
outcomes when compared to fluctuating or short-term payments.

Several federal government programs provide incentives to farmers over
long time periods. For example, the US Department of Agriculture (USDA)
manages a Conservation Stewardship Program (CSP) which is a 5-year
contract -- potentially renewed for 5 more years -- that pays farmers an
annual sum in return for agreeing to implement a customized conservation
plan co-created with a USDA agent. The plan allows growers to build on
their existing conservation practices by implementing practices that
improve a wide range of on-farm conditions, from soils to biodiversity.
USDA also manages the Environmental Quality Improvement Program (EQIP),
which similarly supports on-farm diversification practices with
contracts that typically last 1-3 years but may extend to 10 years.
Payment rates are reviewed and changed annually; certain practices may
receive sizable assistance but rates can be unstable over time
(Agriculture 2021). While both CSP and EQIP are heavily in demand by
farmers in many states, including California (Wyant 2020), researchers
have not yet examined whether the differing longevity of the incentives
provided via these programs could impact the durability of
diversification practice implementation.

In conclusion, by combining semi-structured interview data with a
modeling approach that captures complex temporal dynamics in a stylized
social-ecological system model, we offer insights into important
agricultural management patterns and their implications for ecological
outcomes and public policy. While tipping points have been extensively
studied throughout the social-ecological systems literature, including
agriculture, we suggest a novel mechanism for these tipping points that
makes minimal assumptions about system-specific behavior. Further, we
present a flexible model framework that can be built on to address
critical questions in social-ecological systems research and policy
design.

\hypertarget{experimental-procedures}{%
\section{Experimental Procedures}\label{experimental-procedures}}

\hypertarget{conceptual-model-description}{%
\subsection{Conceptual model
description}\label{conceptual-model-description}}

We explore the transition to and from diversified farming systems (low
and high ecosystem service provisioning states) using a Markov Decision
Process (MDP) in which a farmer makes a series of decisions about
whether or not to employ agricultural diversification practices over
time (Figure 1). In the context of diversified farming systems,
diversification practices include hedgerows, crop rotation, intercrops,
the use of compost, growing multiple crop types, reduced tillage, and
cover crops. These practices have often been shown to promote ecosystem
services that benefit farmers, including soil fertility and
water-holding capacity, pest and disease control, pollination and
productivity, thus providing an economically-viable alternative to
chemically-intensive methods of crop production (Tamburini et al. 2020;
Kremen and Miles 2012; Rosa-Schleich et al. 2019; Beillouin et al.
2021). This type of diversification is distinct from the concept of
operational diversification (i.e., increasing the range of revenue
streams produced on a given farm, such as tourism or value-added
products). The model was developed through an iterative, collaborative
process with an interdisciplinary team comprising plant and soil
scientists, agricultural economists, ecologists, agricultural
sociologists, modelers, policy analysts, and farmers with the goal of
capturing patterns stemming from the coupled human and natural dynamics
of the modeled system.

\hypertarget{conceptual-model-description-1}{%
\subsection{Conceptual model
description}\label{conceptual-model-description-1}}

We explore the transition to and from diversified farming systems (low
and high ecosystem service provisioning states) using a Markov Decision
Process (MDP) in which a farmer makes a series of decisions about
whether or not to employ agricultural diversification practices over
time (Figure 6). Modeling the adoption of diversification practices and
the resultant ecosystem services as a Markov Decision Process requires
that we first define a set of available ``actions'' (or decisions) and a
set of possible system states. In our model, at each time step, the
farmer takes an ``action'' of 0\% to 100\% investment in adopting or
maintaining diversification practices. The ``system state'' corresponds
to the level of benefit derived from the ecosystem services that result
from those adoption decisions. While higher ecological states are
beneficial, investments in diversification practices also come with
higher associated costs (Figure 6 A1). Costs and benefits may be
financial, social, ideological, and/or aesthetic, and we approximate
that relationship as linear (Figure 6 A2). A greater percent investment
in diversification practices corresponds to a greater probability of
transitioning to a higher (more beneficial) ecological state within the
next decision cycle (Figure 6 A3). Our model makes minimal assumptions
about the relationships between actions and costs (Figure 6 A1), states
and benefits (Figure 6 A2), and actions and state changes (Figure 6 A3).
While additional assumptions could be integrated into this MDP framework
(e.g., nonlinear functions for Figure 6 A1-A3), we focus our study on
the impact of the interactions between ecological rates and time
horizons of decisions by minimizing assumptions around the functional
forms of these subsystems. The rate at which that ecological response
occurs depends on parameter r, but importantly is not instantaneous
(Figure 6 A4). By defining parameter values for cost, benefit,
transition stochasticity, ecological change rate, and future
discounting, a Markov Decision Process allows the optimal action
strategy for the farmer (agent) to emerge based on expected rewards
(benefits minus costs) over either a finite (to represent short-tenure
leased farms) or infinite (to represent longer-term leases and land
ownership) time horizon (Figure 6 A5). We use a ten-year time horizon to
represent shorter term decision-making, essentially the longest frame of
reference that tenant farmers tend to work within and a conservative way
of looking at the impact of lease length for tenant farmers (Bigelow,
Borchers, and Hubbs 2016). This frame of reference is suggested not only
in the agrarian sociology literature but in the farmer interviews we
conducted. Discounted infinite decision horizons are meant to represent
landowners and other farmers with the capacity to account for the
economic viability of an action over the long run.

\begin{figure}[H]

{\centering \includegraphics[width=1\linewidth]{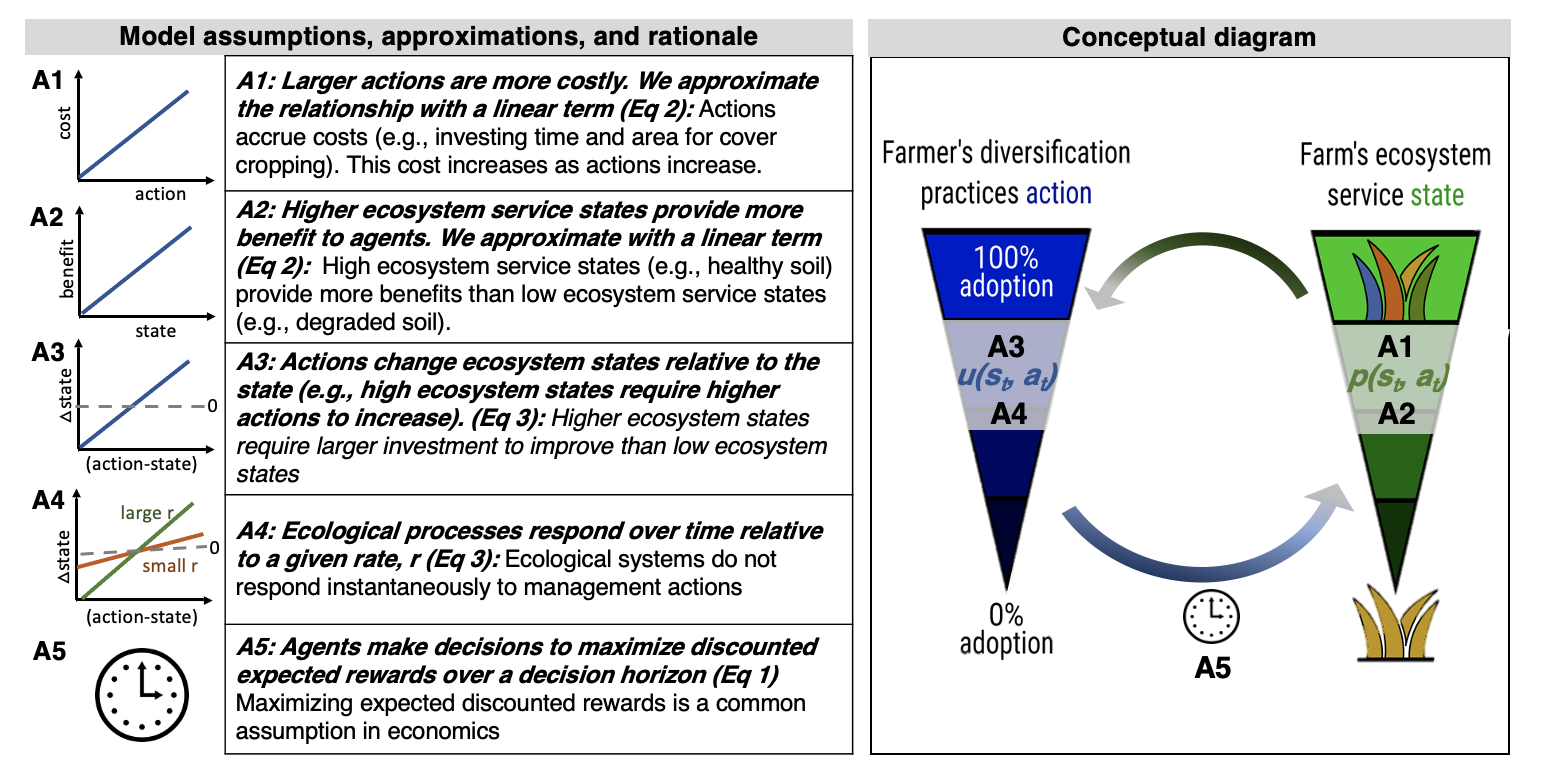} 

}

\caption{Conceptual diagram and model assumptions. The farmer’s choice of how much to invest (time and money) into the adoption of diversification practices is shown in blue, and the resulting ecosystem services state in green, with a more diversified ecosystem state at the top, and a more simplified ecosystem state at bottom. Each time step, the farmer chooses the optimal action for their current ecosystem service state based on the perceived utility function, u, and state transition probability function, p.  For a given ecosystem service state and action at time t, p describes how the ecosystem responds stochastically to result in an updated state at t + 1. The updated ecosystem service state then feeds back to influence the farmer’s future choices, leading to tradeoffs arising from the coupling of ecological processes with consecutive diversification practice adoption decisions over time. Main model assumptions (A1-A5) are outlined along with a brief rationale for each approximation.}\label{fig:assumption}
\end{figure}

\hypertarget{interview-data}{%
\subsection{Interview data}\label{interview-data}}

As part of the larger project that our modeling work contributes to,
between February 2018 - August 2020, the agricultural sociologists in
our team interviewed 25 lettuce growers and 17 almond growers from
California using a snowball sampling method. We developed an interview
guide with questions that focused on the barriers and motivations for
using diversification practices such as cover cropping, planting
hedgerows, and diverse crop rotations. We focused on almonds and leafy
greens/lettuce because these are among the most economically valuable
and regionally prevalent crops in California, represent different
farming systems and environmental conditions, and their increased
diversification could have major impacts (for almonds, a very large
acreage could benefit; for leafy greens, their requirements for
fertilizer and pesticide applications could be reduced greatly). We
selected interviewees to represent a range of growers (small to large
scale; organic to conventional; early adopters of diversification
practices to late adopters; family run to corporate management; and
direct-to-consumer marketing to wholesale). Interviews were conducted in
person or over the phone when in-person interviews were not possible due
to farmer schedules or the need to social distance during COVID-19
restrictions. Most interviews were audio-recorded and transcribed. If
recording was not possible, careful notes were taken to create a
transcript.

We performed deductive coding for central themes and keywords of the
transcripts to inform structural attributes of our model. Specifically,
our interview coding informed the relationships among costs, benefits
and actions in diversified farming systems, the integration of time
horizons into decision strategies, and the gradual rate of ecological
change in response to management actions (Figure 7). Additionally,
interviews provided quotes to contextualize model findings.

\begin{figure}

{\centering \includegraphics[width=.9\textwidth]{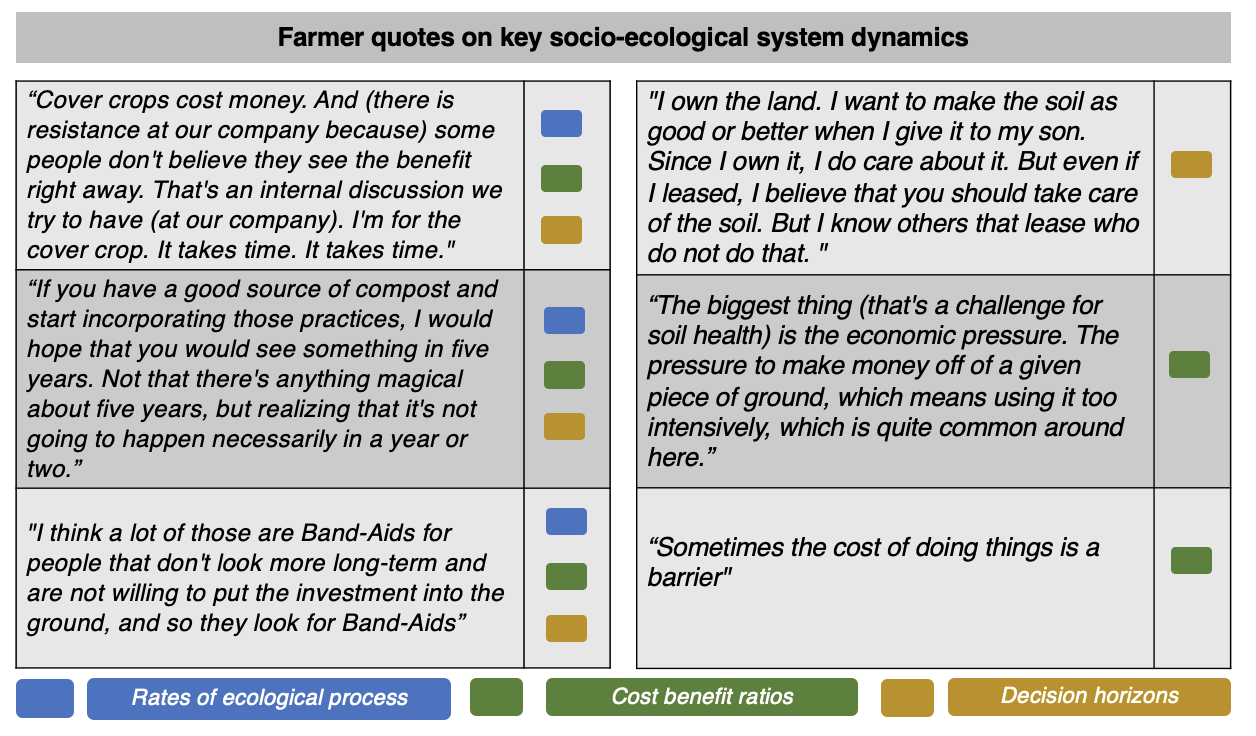} 

}

\caption{Key quotes from farmers suggest that the temporal horizons of decision making and the rate at which farmers receive ecosystem benefits as a result of those decisions are important factors in the adoption of diversification practices}\label{fig:unnamed-chunk-1}
\end{figure}

\hypertarget{mathematical-description}{%
\subsection{Mathematical description}\label{mathematical-description}}

The Markov Decision Process is composed of two coupled models: a model
of the biological/ecological processes, \(s_{t+1} = f(s_t, a_t)\), and a
model of how the farmer views those processes, expressed as the utility
function of the biological state and the cost of the farming
actions/decisions \(u(s_t, a_t)\). Both models incorporate temporal
dynamics. The biological model has a notion of time which says that
actions don't immediately change the biological environment, but instead
change it over time at rate \(r\). Meanwhile, the farmer pays the cost
of action \(a_t\) as soon as that action is taken. However, unlike
common alternative frameworks, such as most agent-based
(`individual-based-simulation') models, the farmer does not choose the
sequence of actions one at a time. Instead, the farmer plans ahead over
the future, by considering actions which may be costly now but pay off
in years to come given the utility of a strategy (i.e.~a sequence of
actions, the discounted sum of the utility of all the individual actions
in the strategy). This decision model can be formulated as

\[\max_{\lbrace a_t \in A\rbrace} \mathbb{E} \left[ \sum_t^T u(s_t, a_t) \gamma^t \right]\]

where \(\lbrace a_t \rbrace\) is chosen from the set of available
actions, \(\mathbb{E}\) the expected utility operator, \(u(s_t, a_t)\)
the utility which the farmer associates with being in state \(s_t\) and
taking action \(a_t\) at time \(t\), \(\gamma\) the myopic discount
factor, and \(T\) the time horizon of the decision, which in this case
represents the land tenure of the farm. In our study we set \(T = 10\)
to represent tenant farms and \(T \to \infty\) to represent a farmer who
owns the land or has a long lease. The farmer takes action \(a_t\) to
get the highest expected return over either an infinite decision horizon
or a given finite decision horizon (methods to solve for the action
policy are outlined in Marescot et al., 2013 (@ Marescot et al. 2013)).

We assume a simple model of the farmer's perceived utility
\(u(s_t, a_t)\) as a function of the difference between the cost \(c_a\)
associated with diversification practice action \(a_t\), versus expected
benefits \(b_s\) derived from ecosystem state \(s_t\), at time \(t\),
such that

\[u(s_t, a_t) = b_s s_t  - c_a a_t\]

where farmers' initial ecosystem states were distributed normally around
a mean of \(s_0 =0.5\). The ecosystem state is also dynamic, evolving
according to the transition probability function \(p(s_t, a_t)\), such
that

\[s_{t+1} = p(s_t, a_t) := s_t + r \left(a_t - s_t \right) + \epsilon\]

where \(\epsilon \sim N(0, \sigma)\). This provides a minimal state
transition model in which the parameter \(r\) sets the natural timescale
at which the ecosystem can respond to changes in land management
decisions, and \(\sigma\) defines the width of the state transition
probability distribution, capturing the noise inherent to ecological
system change.

While we have assumed very basic transition and utility functions for
this stylized model, in general more complicated nonlinear functions for
both the ecosystem state transition and perceived utility could be
substituted into this framework. \# Model implementation

The model was developed in the \emph{R} programming language (R Core
Team 2019). The \emph{MDPtoolbox} library was used to set up and solve
the MDP (Chades et al. 2017). Code for our model and the experiments
conducted in this paper is available freely at
\url{https://github.com/boettiger-lab/dfs-mdp}.

\hypertarget{acknowledgments}{%
\section{Acknowledgments}\label{acknowledgments}}

Funding for this research was provided by the National Science
Foundation grant number CNH-1824871

\hypertarget{author-contributions}{%
\section{Author Contributions:}\label{author-contributions}}

Conceptualization CB, MC, SW, PB, TB, LC, FC, KE, SG, AI, DK, CK, JL,
EO, JO, MR, AS, JT, HW; Data curation: MC, SW, CB; Formal Analysis: MC,
SW, CB; Funding acquisition: TB, AI, CK, DK, CB; Methodology: CB, MC,
SW, PB, TB, LC, FC, KE, AI, DK, CK, EO, JT, HW; Code: MC, SW, CB;
Visualization: MC, SW, CB; Writing -- original draft: MC, SW, CB, LC,
AI; Writing -- review \& editing: CB, MC, SW, PB, TB, LC, FC, KE, SG,
AI, DK, CK, JL, EO, JO, MR, AS, JT, HW

\hypertarget{competing-interests}{%
\section{Competing Interests:}\label{competing-interests}}

The authors declare no competing interests.

\hypertarget{diversity-and-inclusion-statement}{%
\section{Diversity and Inclusion
Statement:}\label{diversity-and-inclusion-statement}}

We worked to ensure ethnic or other types of diversity in the
recruitment of human subjects. We worked to ensure that the study
questionnaires were prepared in an inclusive way. One or more of the
authors of this paper self-identifies as an underrepresented ethnic
minority in science. One or more of the authors of this paper
self-identifies as a member of the LGBTQ+ community. One or more of the
authors of this paper self-identifies as living with a disability.

\hypertarget{references}{%
\section*{References}\label{references}}
\addcontentsline{toc}{section}{References}

\hypertarget{refs}{}
\begin{CSLReferences}{1}{0}
\leavevmode\vadjust pre{\hypertarget{ref-USDApayments}{}}%
Agriculture, United States Department of. 2021. {``2021 State Payment
Schedules.''}

\leavevmode\vadjust pre{\hypertarget{ref-Andow1989}{}}%
Andow, David A., and Kazumasa Hidaka. 1989. {``{Experimental natural
history of sustainable agriculture: syndromes of production}.''}
\emph{Agriculture, Ecosystems and Environment}.
\url{https://doi.org/10.1016/0167-8809(89)90105-9}.

\leavevmode\vadjust pre{\hypertarget{ref-Batary2015a}{}}%
Batáry, Péter, Lynn V. Dicks, David Kleijn, and William J. Sutherland.
2015. {``{The role of agri-environment schemes in conservation and
environmental management}.''} \emph{Conservation Biology}.
\url{https://doi.org/10.1111/cobi.12536}.

\leavevmode\vadjust pre{\hypertarget{ref-beillouin2021positive}{}}%
Beillouin, Damien, Tamara Ben-Ari, Eric Malezieux, Verena Seufert, and
David Makowski. 2021. {``Positive but Variable Effects of Crop
Diversification on Biodiversity and Ecosystem Services.''} \emph{Global
Change Biology}.

\leavevmode\vadjust pre{\hypertarget{ref-bellman1957markovian}{}}%
Bellman, Richard. 1957. {``A Markovian Decision Process.''}
\emph{Indiana Univ. Math. J.} 6: 679--84.

\leavevmode\vadjust pre{\hypertarget{ref-bigelow2016us}{}}%
Bigelow, Daniel, Allison Borchers, and Todd Hubbs. 2016. {``US Farmland
Ownership, Tenure, and Transfer.''}

\leavevmode\vadjust pre{\hypertarget{ref-calo2018knowledge}{}}%
Calo, Adam. 2018. {``How Knowledge Deficit Interventions Fail to Resolve
Beginning Farmer Challenges.''} \emph{Agriculture and Human Values} 35
(2): 367--81.

\leavevmode\vadjust pre{\hypertarget{ref-calo2016after}{}}%
Calo, Adam, and Kathryn Teigen De Master. 2016. {``After the Incubator:
Factors Impeding Land Access Along the Path from Farmworker to
Proprietor.''} \emph{Journal of Agriculture, Food Systems, and Community
Development} 6 (2): 111--27.

\leavevmode\vadjust pre{\hypertarget{ref-Chades2017}{}}%
Chades, Iadine, Guillaume Chapron, Marie-Josee Cros, Frederick Garcia,
and Regis Sabbadin. 2017. \emph{MDPtoolbox: Markov Decision Processes
Toolbox}. \url{https://CRAN.R-project.org/package=MDPtoolbox}.

\leavevmode\vadjust pre{\hypertarget{ref-claassen2014additionality}{}}%
Claassen, Roger, John Horowitz, Eric Duquette, and Kohei Ueda. 2014.
{``Additionality in US Agricultural Conservation and Regulatory Offset
Programs.''} \emph{USDA-ERS Economic Research Report}, no. 170.

\leavevmode\vadjust pre{\hypertarget{ref-Cumming2006a}{}}%
Cumming, Graeme S., David H. M. Cumming, and Charles L. Redman. 2006.
{``{Scale mismatches in social-ecological systems: Causes, consequences,
and solutions}.''} \emph{Ecology and Society}.
\url{https://doi.org/10.5751/ES-01569-110114}.

\leavevmode\vadjust pre{\hypertarget{ref-Dai2012b}{}}%
Dai, Lei, Daan Vorselen, Kirill S. Korolev, and Jeff Gore. 2012.
{``{Generic indicators for loss of resilience before a tipping point
leading to population collapse}.''} \emph{Science}.
\url{https://doi.org/10.1126/science.1219805}.

\leavevmode\vadjust pre{\hypertarget{ref-dayer2018private}{}}%
Dayer, Ashley A, Seth H Lutter, Kristin A Sesser, Catherine M Hickey,
and Thomas Gardali. 2018. {``Private Landowner Conservation Behavior
Following Participation in Voluntary Incentive Programs: Recommendations
to Facilitate Behavioral Persistence.''} \emph{Conservation Letters} 11
(2): e12394.

\leavevmode\vadjust pre{\hypertarget{ref-foley2005global}{}}%
Foley, Jonathan A, Ruth DeFries, Gregory P Asner, Carol Barford, Gordon
Bonan, Stephen R Carpenter, F Stuart Chapin, et al. 2005. {``Global
Consequences of Land Use.''} \emph{Science} 309 (5734): 570--74.

\leavevmode\vadjust pre{\hypertarget{ref-foley2011solutions}{}}%
Foley, Jonathan A, Navin Ramankutty, Kate A Brauman, Emily S Cassidy,
James S Gerber, Matt Johnston, Nathaniel D Mueller, et al. 2011.
{``Solutions for a Cultivated Planet.''} \emph{Nature} 478 (7369):
337--42.

\leavevmode\vadjust pre{\hypertarget{ref-fraser2004land}{}}%
Fraser, Evan DG. 2004. {``Land Tenure and Agricultural Management: Soil
Conservation on Rented and Owned Fields in Southwest British
Columbia.''} \emph{Agriculture and Human Values} 21 (1): 73--79.

\leavevmode\vadjust pre{\hypertarget{ref-gladwell2006tipping}{}}%
Gladwell, Malcolm. 2006. \emph{The Tipping Point: How Little Things Can
Make a Big Difference}. Little, Brown.

\leavevmode\vadjust pre{\hypertarget{ref-gonthier2019bird}{}}%
Gonthier, David J, Amber R Sciligo, Daniel S Karp, Adrian Lu, Karina
Garcia, Gila Juarez, Taiki Chiba, Sasha Gennet, and Claire Kremen. 2019.
{``Bird Services and Disservices to Strawberry Farming in Californian
Agricultural Landscapes.''} \emph{Journal of Applied Ecology} 56 (8):
1948--59.

\leavevmode\vadjust pre{\hypertarget{ref-graddy2017supply}{}}%
Graddy-Lovelace, Garrett, and Adam Diamond. 2017. {``From Supply
Management to Agricultural Subsidies---and Back Again? The US Farm Bill
\& Agrarian (in) Viability.''} \emph{Journal of Rural Studies} 50:
70--83.

\leavevmode\vadjust pre{\hypertarget{ref-Horan2011a}{}}%
Horan, Richard D., Eli P. Fenichel, Kevin L. S. Drury, and David M.
Lodge. 2011a. {``{Managing ecological thresholds in coupled
environmental-human systems}.''} \emph{Proceedings of the National
Academy of Sciences of the United States of America}.
\url{https://doi.org/10.1073/pnas.1005431108}.

\leavevmode\vadjust pre{\hypertarget{ref-Horan7333}{}}%
---------. 2011b. {``Managing Ecological Thresholds in Coupled
Environmental{\textendash}human Systems.''} \emph{Proceedings of the
National Academy of Sciences} 108 (18): 7333--38.
\url{https://doi.org/10.1073/pnas.1005431108}.

\leavevmode\vadjust pre{\hypertarget{ref-Kremen2012}{}}%
Kremen, Claire, Alastair Iles, and Christopher Bacon. 2012.
{``{Diversified farming systems: An agroecological, systems-based
alternative to modern industrial agriculture}.''} \emph{Ecology and
Society}. \url{https://doi.org/10.5751/ES-05103-170444}.

\leavevmode\vadjust pre{\hypertarget{ref-kremen2012ecosystem}{}}%
Kremen, Claire, and Albie Miles. 2012. {``Ecosystem Services in
Biologically Diversified Versus Conventional Farming Systems: Benefits,
Externalities, and Trade-Offs.''} \emph{Ecology and Society} 17 (4).

\leavevmode\vadjust pre{\hypertarget{ref-Lippe2019}{}}%
Lippe, Melvin, Mike Bithell, Nick Gotts, Davide Natalini, Peter
Barbrook-Johnson, Carlo Giupponi, Mareen Hallier, et al. 2019. {``{Using
agent-based modelling to simulate social-ecological systems across
scales}.''} \emph{GeoInformatica}.
\url{https://doi.org/10.1007/s10707-018-00337-8}.

\leavevmode\vadjust pre{\hypertarget{ref-liu2007complexity}{}}%
Liu, Jianguo, Thomas Dietz, Stephen R Carpenter, Marina Alberti, Carl
Folke, Emilio Moran, Alice N Pell, et al. 2007. {``Complexity of Coupled
Human and Natural Systems.''} \emph{Science} 317 (5844): 1513--16.

\leavevmode\vadjust pre{\hypertarget{ref-long2017hedgerow}{}}%
Long, R, Kelly Garbach, L Morandin, et al. 2017. {``Hedgerow Benefits
Align with Food Production and Sustainability Goals.''} \emph{California
Agriculture} 71 (3): 117--19.

\leavevmode\vadjust pre{\hypertarget{ref-Marescot2013b}{}}%
Marescot, Lucile, Guillaume Chapron, Iadine Chadès, Paul L. Fackler,
Christophe Duchamp, Eric Marboutin, and Olivier Gimenez. 2013.
{``{Complex decisions made simple: A primer on stochastic dynamic
programming}.''} \url{https://doi.org/10.1111/2041-210X.12082}.

\leavevmode\vadjust pre{\hypertarget{ref-minkoff2019new}{}}%
Minkoff-Zern, Laura-Anne. 2019. \emph{The New American Farmer:
Immigration, Race, and the Struggle for Sustainability}. MIT Press.

\leavevmode\vadjust pre{\hypertarget{ref-Mumby2007}{}}%
Mumby, Peter J., Alan Hastings, and Helen J. Edwards. 2007.
{``{Thresholds and the resilience of Caribbean coral reefs}.''}
\emph{Nature}. \url{https://doi.org/10.1038/nature06252}.

\leavevmode\vadjust pre{\hypertarget{ref-olimpi2019evolving}{}}%
Olimpi, Elissa M, Patrick Baur, Alejandra Echeverri, David Gonthier,
Daniel S Karp, Claire Kremen, Amber Sciligo, and Kathryn T De Master.
2019. {``Evolving Food Safety Pressures in California's Central Coast
Region.''} \emph{Frontiers in Sustainable Food Systems} 3: 102.

\leavevmode\vadjust pre{\hypertarget{ref-olimpi2020shifts}{}}%
Olimpi, EM, K Garcia, DJ Gonthier, KT De Master, A Echeverri, C Kremen,
AR Sciligo, WE Snyder, EE Wilson-Rankin, and DS Karp. 2020. {``Shifts in
Species Interactions and Farming Contexts Mediate Net Effects of Birds
in Agroecosystems.''} \emph{Ecological Applications} 30 (5): e02115.

\leavevmode\vadjust pre{\hypertarget{ref-Ong2020a}{}}%
Ong, Theresa Wei Ying, and Wenying Liao. 2020. {``{Agroecological
Transitions: A Mathematical Perspective on a Transdisciplinary
Problem}.''} \url{https://doi.org/10.3389/fsufs.2020.00091}.

\leavevmode\vadjust pre{\hypertarget{ref-Ostrom2009}{}}%
Ostrom, Elinor. 2009. {``{A general framework for analyzing
sustainability of social-ecological systems}.''}
\url{https://doi.org/10.1126/science.1172133}.

\leavevmode\vadjust pre{\hypertarget{ref-poeplau2015carbon}{}}%
Poeplau, Christopher, and Axel Don. 2015. {``Carbon Sequestration in
Agricultural Soils via Cultivation of Cover Crops--a Meta-Analysis.''}
\emph{Agriculture, Ecosystems \& Environment} 200: 33--41.

\leavevmode\vadjust pre{\hypertarget{ref-RCT2019}{}}%
R Core Team. 2019. \emph{R: A Language and Environment for Statistical
Computing}. Vienna, Austria: R Foundation for Statistical Computing.
\url{https://www.R-project.org/}.

\leavevmode\vadjust pre{\hypertarget{ref-richardson2015land}{}}%
Richardson Jr, Jesse J. 2015. {``Land Tenure and Sustainable
Agriculture.''} \emph{Tex. A\&M L. Rev.} 3: 799.

\leavevmode\vadjust pre{\hypertarget{ref-roberts2007enduring}{}}%
Roberts, Michael J, and Ruben N Lubowski. 2007. {``Enduring Impacts of
Land Retirement Policies: Evidence from the Conservation Reserve
Program.''} \emph{Land Economics} 83 (4): 516--38.

\leavevmode\vadjust pre{\hypertarget{ref-rockstrom2009safe}{}}%
Rockström, Johan, Will Steffen, Kevin Noone, Åsa Persson, F Stuart
Chapin, Eric F Lambin, Timothy M Lenton, et al. 2009. {``A Safe
Operating Space for Humanity.''} \emph{Nature} 461 (7263): 472--75.

\leavevmode\vadjust pre{\hypertarget{ref-rosa2019ecological}{}}%
Rosa-Schleich, Julia, Jacqueline Loos, Oliver Mußhoff, and Teja
Tscharntke. 2019. {``Ecological-Economic Trade-Offs of Diversified
Farming Systems--a Review.''} \emph{Ecological Economics} 160: 251--63.

\leavevmode\vadjust pre{\hypertarget{ref-Scheffer2010a}{}}%
Scheffer, Marten. 2010. {``{Foreseeing tipping points}.''}
\emph{Nature}. \url{https://doi.org/10.1038/467411a}.

\leavevmode\vadjust pre{\hypertarget{ref-soule2000land}{}}%
Soule, Meredith J, Abebayehu Tegene, and Keith D Wiebe. 2000. {``Land
Tenure and the Adoption of Conservation Practices.''} \emph{American
Journal of Agricultural Economics} 82 (4): 993--1005.

\leavevmode\vadjust pre{\hypertarget{ref-stoate2009ecological}{}}%
Stoate, C, A Báldi, Pl Beja, ND Boatman, I Herzon, A Van Doorn, GR De
Snoo, L Rakosy, and C Ramwell. 2009. {``Ecological Impacts of Early 21st
Century Agricultural Change in Europe--a Review.''} \emph{Journal of
Environmental Management} 91 (1): 22--46.

\leavevmode\vadjust pre{\hypertarget{ref-tamburini2020agricultural}{}}%
Tamburini, Giovanni, Riccardo Bommarco, Thomas Cherico Wanger, Claire
Kremen, Marcel GA van der Heijden, Matt Liebman, and Sara Hallin. 2020.
{``Agricultural Diversification Promotes Multiple Ecosystem Services
Without Compromising Yield.''} \emph{Science Advances} 6 (45): eaba1715.

\leavevmode\vadjust pre{\hypertarget{ref-Vandermeer1997}{}}%
Vandermeer, John. 1997a. {``{Syndromes of production: An emergent
property of simple agroecosystem dynamics}.''} \emph{Journal of
Environmental Management}. \url{https://doi.org/10.1006/jema.1997.0128}.

\leavevmode\vadjust pre{\hypertarget{ref-vandermeer1997syndromes}{}}%
---------. 1997b. {``Syndromes of Production: An Emergent Property of
Simple Agroecosystem Dynamics.''} \emph{Journal of Environmental
Management} 51 (1): 59--72.

\leavevmode\vadjust pre{\hypertarget{ref-Vandermeer2012}{}}%
Vandermeer, John H., and Ivette Perfecto. 2012. {``{Syndromes of
production in agriculture: Prospects for social-ecological regime
change}.''} \emph{Ecology and Society}.
\url{https://doi.org/10.5751/ES-04813-170439}.

\leavevmode\vadjust pre{\hypertarget{ref-vandermeer2012syndromes}{}}%
Vandermeer, John, and Ivette Perfecto. 2012. {``Syndromes of Production
in Agriculture: Prospects for Social-Ecological Regime Change.''}
\emph{Ecology and Society} 17 (4).

\leavevmode\vadjust pre{\hypertarget{ref-Walker2004}{}}%
Walker, Brian, C. S. Holling, Stephen R. Carpenter, and Ann Kinzig.
2004. {``{Resilience, adaptability and transformability in
social-ecological systems}.''} \emph{Ecology and Society}.
\url{https://doi.org/10.5751/ES-00650-090205}.

\leavevmode\vadjust pre{\hypertarget{ref-wyant_2020}{}}%
Wyant, Sara. 2020. {``A Closer Look at EQIP - One of NRCS's Most Popular
Conservation Programs.''} \emph{Agweek}. Agweek.
\url{https://www.agweek.com/opinion/columns/6784838-A-closer-look-at-EQIP-\%E2\%80\%94-one-of-NRCS\%E2\%80\%99s-most-popular-conservation-programs}.

\end{CSLReferences}

\bibliographystyle{unsrt}
\bibliography{references.bib}

\end{document}